\documentclass{optica-article}

\journal{opticajournal}

\articletype{Research Article}

\usepackage{lineno}
\usepackage{orcidlink}
\newtheorem{theorem}{Theorem}
\newtheorem{corollary}[theorem]{Corollary}
\newtheorem{case}[theorem]{Case}
\newtheorem{definition}[theorem]{Definition}
\newtheorem{conclusion}[theorem]{Conclusion}
\begin{document}

\title{The Optics of Shadow Bands}

\author{Branko Sretenović\orcidlink{0009-0008-3175-3781}\authormark{1,*}}

\address{\authormark{1}Independent researcher, Serbia}

\email{\authormark{*}mathallica3@outlook.com}

\begin{abstract*}
Shadow bands are transient, rippling patterns of light and dark that may appear in moments before and after totality in a solar eclipse. Despite centuries of reports, their physical origin has remained unresolved. This article develops a geometric-optical solution in which the Sun's extended structure produces a celestial analog of Young's double-slit experiment, generating an interference and fringes on the ground. The analysis combines solar limb geometry and a wave-based formulation that yields quantitative predictions for fringe width and spacing. Some other properties of shadow bands are also analyzed.
\end{abstract*}

\section{Introduction}
Shadow bands are transient, centimeter-scale fluctuations of light and dark that appear on the ground in the moments before and after totality during a solar eclipse. Although reported for more than two centuries, their physical origin remains unresolved. Observations show variability in visibility, spacing, motion, and contrast, and existing explanations account for only subsets of these features. No quantitative predictive framework has been established.

Previous attempts, including atmospheric-turbulence models such as Codona's, reproduce certain qualitative aspects but do not yield quantitative predictions for fringe spacing or directional motion, nor do they identify the underlying physical cause in the solar-atmospheric system. These limitations motivate the geometric-optical model based on solar-limb geometry and wave-based propagation developed in the sections that follow.

Similar banded patterns have occasionally been reported during other obscuration events, such as sunrise and sunset, suggesting a broader underlying mechanism.

\section{Methods}
The aim of this work is to identify the physical mechanism responsible for the shadow bands and to derive approximate predictions for their observable width, spacing, and motion. The analysis is theoretical and draws on established solar-limb geometry, optical principles, and documented observational constraints.

Atmospheric variability and solar-atmospheric irregularities can influence the appearance of shadow bands, introducing uncertainties that are not modeled in detail here. The focus is instead on the underlying geometric-optical mechanism that governs the formation and evolution of the bands under clear terrestrial conditions.

To establish that the physical analog of shadow bands is the system of bright-dark fringes produced in a double-slit experiment, we treat the two bright solar shafts present at the verge of totality as the effective pair of coherent sources. Although the modern double-slit experiment is typically performed with a laser illuminating two slits in a darkened laboratory, such controlled conditions are not essential for the underlying physics. Young's original demonstration---achieved with nothing more than a pane of glass and a thin card---already shows that a two-slit configuration can arise from simple, naturally occurring geometries.

For a double-slit interference pattern to form, several physical conditions must be met. The light must possess sufficient coherence to maintain a stable phase relationship across the two paths. Monochromatic, or at least quasi-monochromatic, illumination is strongly preferred, as a narrow spectral bandwidth preserves the phase stability required for high-contrast fringes. The apertures must be narrow, well-defined, and separated by a distance that allows their diffraction patterns to overlap on the observation plane. The screen must be far enough from the slits for the diffracted waves to spread and interfere, and the environment must be sufficiently stable and darkened to preserve fringe contrast. Under these preconditions---even with minimal equipment, as in Young's demonstration---the characteristic bright-dark fringe system emerges with clarity.

Beyond establishing the analogy between shadow bands and double-slit fringes, the model must also satisfy additional observational constraints, including the temporal evolution, directional motion, altitude dependence, and sensitivity to surface texture reported in eclipse observations. These constraints guide the geometric-optical formulation developed in the sections that follow.

\section{Observational Constraints}
Shadow bands do not always appear during a total solar eclipse. When they do, they emerge at specific time intervals---typically just before totality (C2, \textit{second contact})---before vanishing during totality. They reappear after totality (C3, \textit{third contact}) and disappear again after another interval. Observations suggest varying time intervals. However, they always disappear during the totality.

In some cases, shadow bands disappear, and in other cases, they reappear. Sometimes they appear as faint parallel alternating light and dark bands, and other times as curved, snake-like patterns. The recorded widths range from approximately $2.5\,\mathrm{cm}$ ($1\,\mathrm{in}$) to at least $46\,\mathrm{cm}$ ($18\,\mathrm{in}$), while their observed speed varies widely, from approximately $10\,\mathrm{cm/s}$ ($4\,\mathrm{in/s}$) to $180\,\mathrm{cm/s}$ ($70\,\mathrm{in/s}$). Their direction appears tangential to the Moon's shadow. Observers have reported occasional variations in contrast.

Furthermore, these properties evolve as the Moon progressively obscures the solar disk. The visibility and contrast of shadow bands generally increase as the solar crescent narrows.

To ensure an exact and reliable understanding of shadow bands and their effect, the text will incorporate quotations and rephrased findings from authors whose research is widely recognized in the field. Additional references to reputable scientific papers will support and enrich the analysis with valuable insights.

Hence, these observations will be treated as reliable empirical facts regarding shadow bands. The observed facts are categorized according to the distinct properties of shadow bands and will later be compared with the corollaries derived from the proposed explanation to validate its predictions.

\subsection{Width and Spacing}
J. L. Codona developed the scintillation theory of shadow bands, which attributes their formation to atmospheric turbulence acting on light from the solar crescent. In his \textsl{Sky and Telescope} article \emph{The Enigma of Shadow Bands} Codona (1991) presented a plot with time on the $x$-axis, and shadow band spacing on the $y$-axis, illustrating how the shadow-band spacing changes as totality approaches \cite{Codona1991}. From the plot, we can infer that three minutes before totality, the shadow band spacing typically ranges from a few to about $10$ centimeters. The range gradually shrinks, and by one minute before totality the spacing ranges from approximately two to a few centimeters. In the final $20$ seconds, the spacing increases more, with the pattern becoming intricate and richly detailed. On the verge of totality, the range of spacing shrinks to its minimum, but the spacing somewhat increases and is generally a few centimeters. Shadow bands become more distinctly visible. Nevertheless, the spacing may extend to half a meter, to a meter, or even longer.

In the article \emph{Visual, Photographic and Photoelectric Detection of Shadow Bands at the March 7, 1970, Solar Eclipse}, published in the international journal \textsl{Nature} in 1971, a team from the Department of Physics at Ball State University, Indiana, measured several characteristics of shadow bands, focusing on their orientation, width, spacing, and speed \cite{Hults1971}. Teams of observers at evenly distributed locations around the Moon's shadow on Earth, known as the \textit{umbra,} collected information visually, photographically, and photoelectrically. The width and spacing of the shadow bands varied widely. The reported spacing intervals ranged from $3$ to $8$ centimeters and from $15$ to $20$ centimeters while the individual band widths were typically only a few centimeters. Both width and spacing were observed to change with time.

\subsection{Direction}
Feldman (1938), in his article \emph{Shadow Bands - Part I} from \textsl{Popular Astronomy}, analyzed a substantial number of shadow band observations, focusing specifically on their direction. Here is a quote from his conclusion \cite{Feldman1938}.

\begin{quote}
A candid world must therefore accept my statement that the shadow bands are in every case where carefully observed, parallel to the nearest edge of the shadow and consequently a portion of a complete ring of bands.
\end{quote}

In the article \emph{Visual, Photographic and Photoelectric Detection of Shadow Bands at the March 7, 1970, Solar Eclipse} teams of observers at evenly distributed locations around umbra collected data about orientation of shadow bands. In summary, along the umbra central path of the eclipse (\textit{totality path}), the shadow bands aligned tangentially with the eclipse shadow both before and after totality. However, at locations slightly off the centerline of totality path---yet still within the totality path---the bands were only approximately tangential, and in one instance, they exhibited significant rotation during the one-minute period of shadow band activity before totality.

Shadow bands may change direction, and that was stated in a passage from \textsl{The Total Solar Eclipse, 1905} by British Astronomical Association (1906) \cite{BAA1906}.

\begin{quote}
Three minutes before totality they were going in a direction $22^\circ$ west of south. A minute later they had swung round and were only $17^\circ$ west of south. This rotatory movement has been noticed before, but it was
fully verified on this occasion.
\end{quote}

\subsection{Related Phenomena}
O'Meara (2009) in his article \emph{Secret Sky - Searching for Shadow Bands} from \textsl{Astronomy} describes his experience with shadow bands occurring when the Sun was not eclipsed by the Moon \cite{OMeara2009}. He recognized shadow bands that occurred when the Sun was partially eclipsed by a roof while reading his book. The shadow bands rippled diagonally over the pages. In the other situation $15$ minutes before sunrise, while in a jet at high altitude, he saw shadow bands on the jet's wing.

In this instance of non-eclipse shadow bands, they appeared soon after sunrise and immediately before sunset. The quote is taken from the article \emph{Observations of ``Shadow Bands'' without an Eclipse}, published in \textsl{Monthly Weather Review 1906} by Henry \cite{Henry1906}.

\begin{quote}
Heretofore this interesting and mysterious phenomenon has been observed only during the occurrence of solar eclipses, but by a very simple method M. Rozet has been able to make daily observations of the bands at sunrise and sunset. The light of the sun at the time of its appearance and disappearance behind somewhat lofty mountains on the horizon is received on a white screen, arranged in the observer's room, and bands are produced apparently identical in character with those observed during an eclipse.
\end{quote}

Shadow bands were artificially generated, observed, and photographed by stars other than the Sun, and these bands appear to share the same underlying cause as shadow bands seen during a total solar eclipse. Here are a few outlines from the Gaviola (1948) article \emph{On Shadow Bands at Total Eclipses of the Sun} in \textsl{Popular Astronomy} \cite{Gaviola1948}.

\begin{quotation}
Shadow bands whose characteristics resemble the ones described are familiar to observing astronomers: they are seen when a bright star is in the field and eyepiece is removed, replacing it with the naked eye. Although easy to see, they are difficult to photograph.

\qquad \lbrack \ldots ]\newline
Picture 1 was taken with light of $\beta$ Centauri using the $12\,\mathrm{cm}$ focus camera and an exposure time of $1/10$ second. Two shadow band system can be seen, having both the same wavelength, $L=7\,\mathrm{cm}$, but different
orientations: $\mathit{S22}\,\mathrm{^\circ}$ \textit{W} and $\mathit{S65}\,\mathrm{^\circ}$ \textit{W}.
\end{quotation}

\subsection{Occurrence}
Observations typically document the time intervals during which shadow bands
appear before totality and fade after it. These periods range from several
seconds to a few minutes.

Feldman (1938) in the article \emph{Shadow Bands - Part I} wrote a paragraph in which he stated that the shadow bands were clearly detected at a height of $3.8$ kilometers during an eclipse \cite{Feldman1938}. The width of the shadow band at the altitude was shorter than on the ground. He commented that the shadow bands were nearly tangential to the umbra.

\begin{quote}
The hearty cooperation of the Spanish government resulted in observations being secured from certain military balloons at Burgos, Spain, from which ``white screens hang horizontally'' \ldots ``In one of the balloons the bands were more clearly seen at the height of $3\,800$ meters than on the ground, and were much finer ($0.6$ to $0.8\,\mathrm{cm}$) and nearer together ($1.2$ to $1.6\,\mathrm{cm}$), and had a velocity of $1.5$ meters per second.'' (On the ground the
widths reported varied from $1$ to $10\,\mathrm{cm}$, and distance apart from $3$ to $25\,\mathrm{cm}$.) The direction reports came from eight different points, through Spain and Majorca, to Tripoli, and Rotch wrote: ``The observations at different places indicate that the bands lay nearly tangent to an ellipse having for its longest diameter the width of the shadow, and representing all points where totality occurred at the same instant.''
\end{quote}

A research team from the Department of Physics and Astronomy and the School of Engineering at the University of Pittsburgh, in collaboration with the Allegheny Observatory, investigated the possible atmospheric origin of shadow bands during the total solar eclipse of August 21, 2017. Their findings were presented in a preprint titled \emph{Observation of Eclipse Shadow Bands Using High Altitude Balloon and Ground-Based Photodiode Arrays} \cite{Madhani2020}. The team deployed photodiode arrays both on the ground and in high-altitude balloons spanning approximately $25\,\mathrm{km}$. Shadow bands were detected in both environments, with continuous and approximately the same fluctuations observed at high altitude and at ground level, suggesting that the phenomenon cannot originate solely within the lower atmosphere of the Earth.

Very few photographs of shadow bands have been found that have not been visually enhanced---typically by increasing contrast. Although shadow bands have been captured on video numerous times, they remain notoriously difficult to photograph. Even when recorded successfully, their features can be elusive and hard to interpret, often leading to confusion. Observers at the same location and moment have sometimes disagreed on what they witnessed.

However, many unaltered shadow band videos are available online. Their motion, once in play, eventually makes them discernible.

\subsection{Shadow Band Theories}
Let us now consider the most widely accepted explanation of shadow bands to date: Scintillation theory. This hypothesis was proposed by Codona (1986), in a peer-reviewed article titled \emph{The Scintillation Theory of Eclipse Shadow Bands}, published in \textsl{Astronomy and Astrophysics} \cite{Codona1986}.

Codona's theory suggests that minute atmospheric irregularities cause light from the solar crescent to undergo scintillation, producing interference-like patterns of alternating light and dark bands on the ground just before and after totality. Although the theory remains prominent, it is not without its critics.

Nevertheless, certain observational findings raise questions regarding the completeness of this explanation. Two such considerations are discussed below.

\begin{enumerate} \renewcommand{\labelenumi}{(\roman{enumi})}
\item \textbf{Concentric Ring Formation}. Empirical studies from independent researchers suggest that shadow bands may form complete rings---some reportedly concentric around the umbra. However, it is unlikely that atmospheric disturbances consistently result in such an organized motion. Although air cooling may influence wind direction toward or away from the umbra's center, its effect near the edges is likely minimal.
\item \textbf{High-Altitude Observations}. Codona's theory attributes shadow band formation to atmospheric turbulence at altitudes below two kilometers. However, empirical evidence has challenged this assumption. Two independent studies---referenced above---have recorded shadow bands at significantly higher elevations: one at $3.8\,\mathrm{km}$, and another at approximately $25\,\mathrm{km}$. These findings suggest that the phenomenon may not originate only within the lower atmosphere.
\end{enumerate}

\section{Model}
The cause of shadow bands comprises the Sun's layers, Earth's atmosphere, features of the umbra, and the properties of light.

The Sun and its atmosphere consist of several layers. For our investigation, we do not need to examine the layers within the Sun---only the one at its surface, that is, the \textit{Photosphere}.

The Sun's limb refers to the outer edge of the Sun's disk as observed from Earth. As one observes the Sun gradually approaching the limb, it appears darker.

The Sun's atmosphere consists of four layers: the \textit{Temperature Minimum Region}, \textit{Chromosphere}, \textit{Transition Region}, and \textit{Corona}. The Temperature Minimum Region is the Sun's lowest atmospheric layer, while the Corona is the uppermost.

The Temperature Minimum Region is often considered an integral part of the Chromosphere. However, it is sometimes regarded as a boundary between the Photosphere and the Chromosphere.

This boundary plays an important role in the study of shadow bands. Hence, in this article the Temperature Minimum Region will specifically refer to a boundary---a distinct entity separate from the Chromosphere.

Although the Sun appears uniform to an observer on Earth, its layers are highly dynamic and emit varied illumination. Below is a brief overview of how bright each layer is in the Sun's surface and atmosphere:

\begin{enumerate} \renewcommand{\labelenumi}{(\roman{enumi})}
\item \textbf{Photosphere}. By far the brightest layer. The Sun disk appears darker at the limb due to observational perspective, yet it remains significantly brighter than any atmospheric layer.
\item \textbf{Temperature Minimum Region (Dark Band)}. Darker than its surroundings.
\item \textbf{Chromosphere}. Initially bright, then gradually darkens.
\item \textbf{Transition Region}. The darkest of all layers in visible light.
\item \textbf{Corona}. Begins bright but quickly dims.
\end{enumerate}

From this we infer the presence of three brighter regions: the Sun disk, the Chromosphere ring, and the Corona ring. Since both the Chromosphere and Corona begin with bright regions when moving outward from the Sun's center, we will refer to these zones as the Chromosphere base and Corona base.

There is no proportionality of the Sun layers' widths in the illustration below (Fig. 1); it represents their relative positions and shape only.

\begin{figure}[htbp]
\centerline{\includegraphics[width=0.85\textwidth, keepaspectratio]
{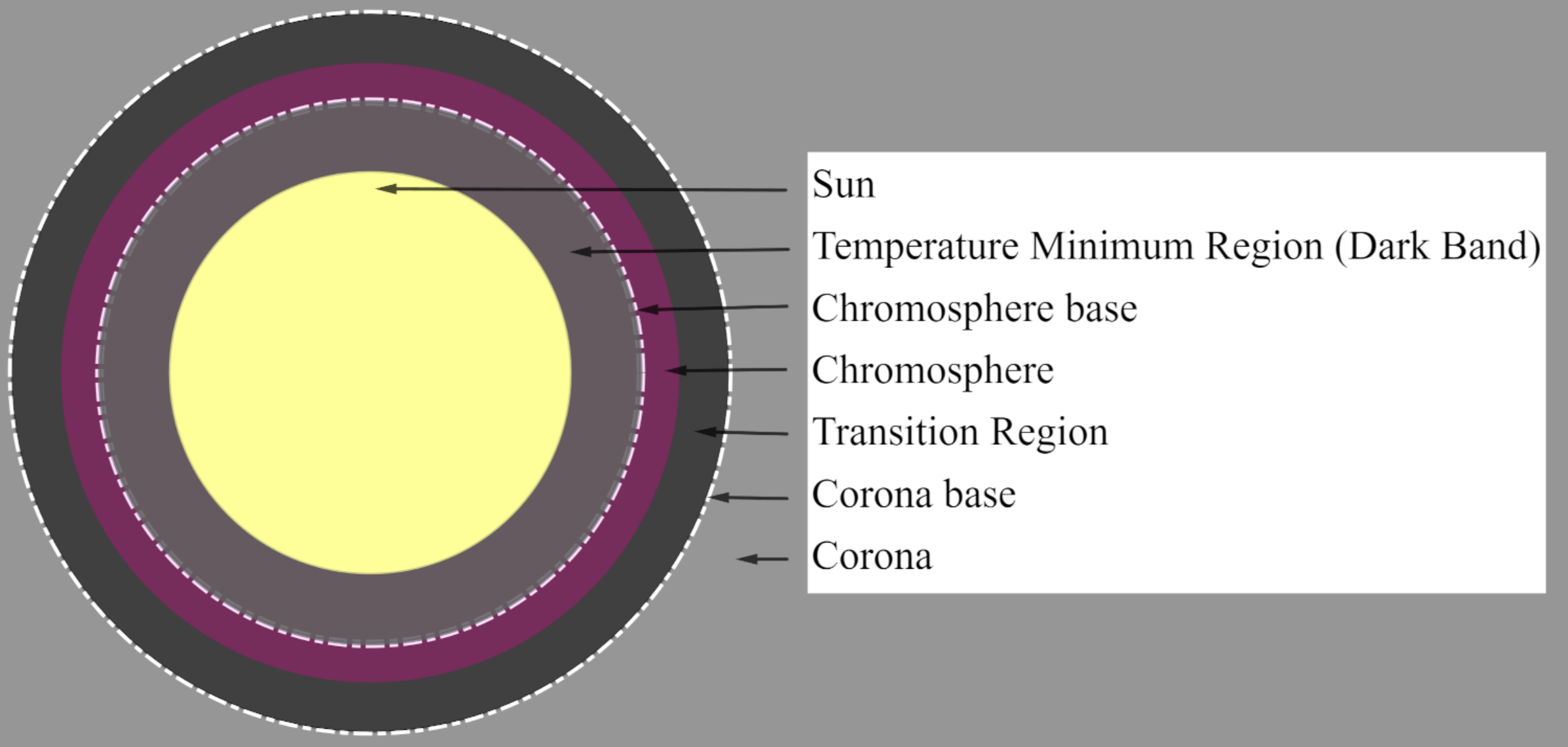}}
\caption{Bright Regions with High Contrast in Sun Layers}
\end{figure}

\noindent Bright strata with high contrast of the Sun and its atmosphere are:

\begin{enumerate}
\item The Sun
\item The Chromosphere base
\item The Corona base
\end{enumerate}

Compared to the Sun's surface, the Sun's atmosphere has a lower intensity and is much less broadband. Each atmospheric layer contributes more specific and darker wavelengths, emphasized in a discrete spectrum.

\subsection{Solar Atmosphere Layer Spectrum}
The Photosphere, the Sun's outermost visible layer, is the primary source of outward-radiating photons. Its light is broadband, covering a full range of visible wavelengths. Due to its intense brightness and spectral richness,
the Photosphere appears visually uniform, effectively concealing the solar atmosphere from direct observation. However, during an eclipse, when the Photosphere is obscured by the Moon, previously hidden layers---such as the
Chromosphere and Corona---briefly emerge into view.

The central region of the Sun's disk appears brighter than its outer edge, a phenomenon known as \textit{limb darkening}. This gradual dimming toward the limb arises from optical and geometric effects.

This effect of \textit{optical depth }is fundamental to how different regions of the Sun manifest during an eclipse, contributing to the layered contrast that defines the solar edge.

Limb darkening is not unique to the Sun; many stars exhibit similar effects. As with our Sun, the intensity of emitted light typically decreases from the stellar center toward the limb due to optical depth and temperature gradients in the outer layers. Simultaneously, the spectral characteristics of stellar atmosphere become more pronounced at the limb of many stars, as they do at the Sun's limb.

Limb darkening is a precondition for the formation of shadow bands.

When observed through optical instruments, the Temperature Minimum Region is often referred to as the \textit{Dark Band}. Since this study focuses on optical phenomena, the term Dark Band will be used frequently throughout the article.

The Dark Band appears darker than the underlying solar surface---but more importantly, it is dimmer than the Chromosphere above. However, the Dark Band should not be perceived as truly dark, but rather as less bright compared to its surroundings.

As will be shown, the Dark Band plays a critical role in producing shadow bands.

The Chromosphere is an extremely dynamic and inhomogeneous layer of the Sun, and it remains only partially understood. It appears progressively darker as it is observed farther from the Sun's center.

Limb darkening enhances the visibility of the Chromosphere near the solar limb, where its spectral lines become more pronounced. However, these lines are not emphasized as strongly as those of the Corona, which lies above the Chromosphere and, therefore, closer to Earth's direct line of sight.

The Chromosphere plays a crucial role in generating shadow bands.

The Transition Region primarily emits ultraviolet light, which lies beyond human visual perception. As such, it appears dark within the visible spectrum---indeed, it is the dimmest layer among all solar strata at visible wavelengths.

The Transition Region---plays a subtle yet significant role in the appearance of shadow bands, but it will not be discussed in this document.

The Corona becomes rapidly darker as we look at it farther from the Sun center.

The Corona lines are emphasized in the Sun limb due to the limb darkening. Its lines are more pronounced than the Chromosphere lines because the Corona is the topmost layer and, therefore, closer to the Earth.

The Chromosphere base and the Corona base are close to the Photosphere. When the Moon partially covers the Photosphere, it also partially covers the Chromosphere base and the Corona base. Consequently, the brightest parts of the Sun surface and atmosphere during an eclipse arise from these three partially covered bases. We'll refer to them as three slivers:

\begin{enumerate}
\item Sun sliver base
\item Chromosphere sliver
\item Corona sliver
\end{enumerate}

The Corona sliver is by far the least bright and we will not analyze its contribution to the shadow band formation.

\subsection{Solar Atmosphere Layer Width}
The extensions of the Sun's atmospheric layers will be presented as perceived from Earth's perspective, referring to them as widths.

Due to solar dynamics, dimensions such as layers' widths cannot always be determined with precision---sometimes carrying uncertainties of several hundred kilometers.

Consequently, the mathematical derivations in this study will often rely on ranges and approximations, reflecting the variable nature of the parameters involved.

While the Photosphere's internal width is not central to this investigation, its visible disk---the apparent surface---defines the Sun's overall size.

We denote the radius of the Sun by \medskip

$R$ --- Sun radius
\begin{eqnarray*}
R &=&7\times 10^{5}\,\mathrm{km} \\
&=&7\cdot 10^{2}\cdot \left( 10^{3}\,\mathrm{km}\right) \\
&=&700\,\mathrm{Mm}
\end{eqnarray*}

The limb zone is not strictly defined, but it is usually taken as the outermost couple percents of the Sun's radius.

The Temperature Minimum Region exhibits significant variability, even across relatively short solar segments. This layer, shaped by the Sun's dynamic behavior, can extend over bright mottles within the Chromosphere---or in some areas between the Sun and Chromosphere, may not manifest at all.

The Dark Band may extend up to $1\,000\,\mathrm{km}$ above the solar surface, or contract to just a few hundred kilometers---if it forms at all.

For our analysis, we adopt an average thickness of $500\,\mathrm{km}$, equivalent to $0.5\,\mathrm{Mm}$.

Estimating the Chromosphere's width is inherently complex due to its dynamic and highly variable nature. Recall that in this article, we distinguish the Chromosphere from the Temperature Minimum Region.

The Chromosphere's brightness increases rapidly from the Dark Band (Temperature Minimum Region). It forms the brightest features at the Chromosphere base.

This stratum typically does not exceed $2\,000\,\mathrm{km}$ ($2\,\mathrm{Mm}$) in altitude. The maximum thickness of the Chromosphere is estimated to not exceed $2\,500\,\mathrm{km}$, although in some regions it may be several times thinner. For the purposes of this study, we adopt a vertical distance between the Temperature Minimum Region and the Chromosphere base of $400\,\mathrm{km}$ ($0.4\,\mathrm{Mm}$).

We estimate the full distance from the Chromosphere base to the solar surface by appending the average thickness of the Dark Band ($0.5\,\mathrm{Mm}$): \medskip

$\delta _{3}$ --- distance from Chromosphere base to Sun surface
\begin{eqnarray*}
\delta _{3} &=&0.5\,\mathrm{Mm}+0.4\,\mathrm{Mm} \\
&=&0.9\,\mathrm{Mm}
\end{eqnarray*}

From the Chromosphere base forward, the brightness drops faster at first but then decreases slowly toward the Transition Region.

The Transition Region, which bridges the Chromosphere and the Corona, is exceptionally dynamic and structurally unstable.

Its thickness varies greatly across the solar disk and within active regions, often measuring only a few hundred kilometers.

The base of the Corona typically begins at an altitude of approximately $2\,500\,\mathrm{km}$ above the Sun's surface (Photosphere), though local variations may shift this by several hundred kilometers.

The brightness of the Corona diminishes rapidly outward from its base, often following an approximate inverse-square law---where intensity decreases with the square of the radial distance from the Sun's center. As a result, the brighter portion of the Corona---referred to as the Corona base---is relatively narrow.

To estimate the distance from the Corona base to the Chromosphere base, we combine known layer altitudes:\medskip

$d_{1}$ --- distance from Corona base to Chromosphere base
\begin{eqnarray*}
d_{1} &=&2\,500\,\mathrm{km}-\delta _{3} \\
&=&2.5\,\mathrm{Mm}-0.9\,\mathrm{Mm} \\
&=&1.6\,\mathrm{Mm}
\end{eqnarray*}

Fig. 2 denotes the Sun and its atmosphere strata in high contrast along with the mutual distances. Notice that the Chromosphere base and the Corona base are represented with a dashed line designating their dark gaps. The picture is descriptive, not drawn to scale.\medskip

\begin{figure}
\centerline{\includegraphics[width=0.60\textwidth, keepaspectratio]
{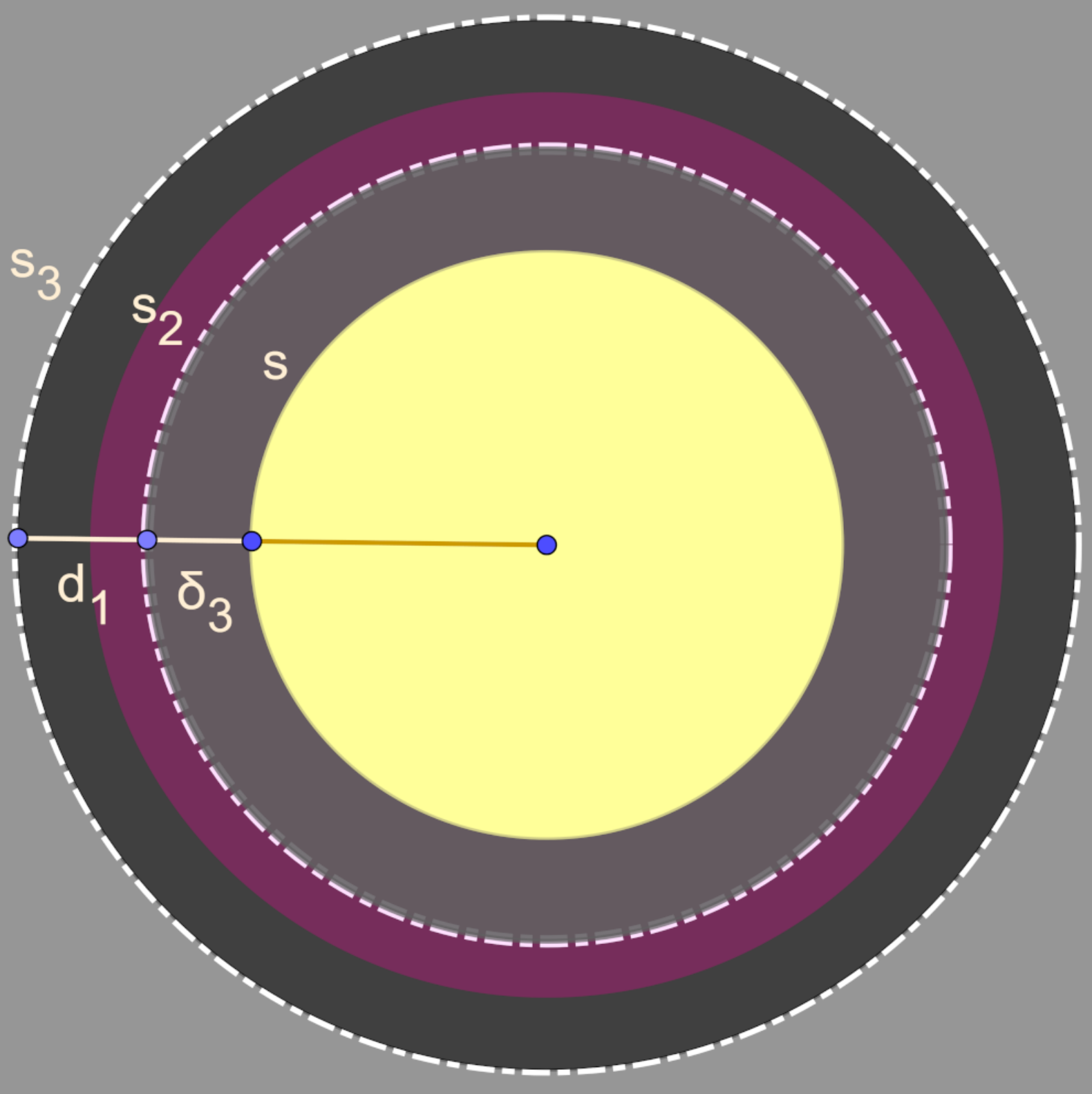}}
\caption{Bright Areas of Sun Surface and Atmosphere with Distance Labels}
\end{figure}

$s$ --- circle representing Sun surface

$s_{2}$ --- circle representing Chromosphere base

$s_{3}$ --- circle representing Corona base\medskip

The distance from the Corona base to the Sun surface is
\begin{eqnarray*}
d_{1}+\delta _{3} &=&0.9\,\mathrm{Mm}+1.6\,\mathrm{Mm} \\
&=&2.5\,\mathrm{Mm}
\end{eqnarray*}

This distance is considerably smaller than the Sun's diameter, that is
\begin{eqnarray*}
2R &=&2\cdot 700\,\mathrm{Mm} \\
&=&1\,400\,\mathrm{Mm}
\end{eqnarray*}

We could say that the Corona and Chromosphere bases are tied to the Sun surface.

\subsection{Geometric Model}

Centuries of shadow band observations have revealed that optimal viewing conditions demand a flat, bright, and non-reflective surface with minimal light dispersion and moderate absorption. Typical examples include white sheets, projection screens, sidewalks, and similar materials.

We refer to any such suitable surface as the \textit{shadow band observation surface}, or briefly, the \textit{observation surface}.

For maximum visibility, the observation surface should be positioned perpendicularly to incoming sunlight rays, allowing the alternating light and dark patterns to form with the greatest clarity.

As a solar eclipse progresses and the Sun is reduced to a thin sliver---only a fraction of an arcminute wide---ambient darkness deepens. Yet even then, the Sun remains Earth's dominant light source.

The darkness of the umbra reveals an elliptical form, shaped by celestial geometry.

As the Moon moves, its umbra travels along the path of totality---the shadow's trajectory across Earth's surface.

The umbra's behavior also depends strongly on the angle of incoming sunlight at the observation surface, meaning both the shape and dynamics of the totality path can vary considerably.

The diameter of the umbra typically ranges from $100\,\mathrm{km}$ ($60\,\mathrm{mi}$) to $1\,000\,\mathrm{km}$ ($600\,\mathrm{mi}$), with a common value of approximately $150\,\mathrm{km}$ ($95\,\mathrm{mi}$).

Near the equator, the lunar umbra travels across Earth at roughly $1\,800\,\mathrm{km}/\mathrm{h}$ ($1\,100\,\mathrm{mi}/\mathrm{h}$), while near the poles, speeds may reach $8\,000\,\mathrm{km}/\mathrm{h}$ ($5\,000\,\mathrm{mi}/\mathrm{h}$). A typical average is around $3\,000\,\mathrm{km}/\mathrm{h}$ ($1\,900\,\mathrm{mi}/\,\mathrm{h}$), although this can vary by a factor of two or more depending on location and geometry.

Light from various solar layers extends outward and grazes the umbra's edge, forming the \textit{penumbra}---a transitional zone where sunlight only partially reaches the surface, and brightness fades gradually.

For analytical clarity, we model the umbra as an ellipse, while acknowledging that subtle adjustments may be needed to account for real-world deviations and temporal changes driven by shifting orbital mechanics.

To illustrate how solar layer illumination affects the observation surface, Figs. 3 and 4 establish key spatial relationships.

The umbra is denoted by \medskip

$u$ --- ellipse representing umbra \medskip

The umbra center lies along the line connecting the centers of the Sun and Moon: \medskip

$C$ --- Sun center

$M$ --- Moon center \medskip

Point $C$ is also the center of the Chromosphere base and the Corona base.

We can interpret the umbra's center as the projection of the Sun's and Moon's centers. \medskip

$C^{\prime }=M^{\prime }$ --- projections of the Sun's and Moon's centers onto the umbra center \medskip

The \textit{centerline} of the umbra is the midline that runs through its core during an eclipse.

From our chosen perspective given in Fig. 4, the umbra moves horizontally, traveling from left to right. The movement direction is represented by a vector originating from the center of the umbra. Therefore, the vector replaces the centerline, visually indicating the umbra's trajectory.

The center point of the observation surface is located near the umbra. \medskip

$O$ --- observation surface center point \medskip

From the observation surface's perspective, it perceives the Sun's and Moon's centers, along with their disks, as in Fig. 3. \medskip

$s$ --- disk representing Sun

$m$ --- disk representing Moon\medskip

Due to limb darkening, the brightest arc of the Sun sliver is the arc bordering the Moon's disk, that represents the Sun sliver base. This arc is the brightest arc during an eclipse, influencing optical effects near totality.

Hence, the Sun sliver brightest point is on the Sun sliver base. Due to limb darkening, the brightest point is the closest point to the Sun center. \medskip

$S_{1}$ --- brightest point of Sun base

$A,\,B$ --- intersections of Sun and Moon disks

$\overset{\frown}{AS_{1}B}$ --- arc of Sun sliver base (brightest arc of Sun sliver) \medskip

Point $S_{1}$ is at the center of the arc $\overset{\frown}{AS_{1}B}$ and is positioned at the width of the Sun sliver. Therefore, it is collinear with the Moon's center $M$ and the Sun's center $C$.

Let us denote the projection of point $O$ onto disks. \medskip

$O^{\prime }$ --- projection of point $O$ onto the plane of the Sun and Moon disks \medskip

Since points $O$, $C^{\prime }$, and $M^{\prime }$ are collinear, points $O^{\prime}$, $C$, $M$, and $S_{1}$ also must be collinear (Fig. 3 and 4).

\begin{corollary}
Projections of the Sun's center, Moon's center, Sun sliver width, and observation surface's center onto the plane of the observation surface are collinear.
\end{corollary}

Point $S_{1}$ is the brightest location on the Sun sliver and the closest point of the Sun sliver base to the observation surface point $O$. It projects light with maximum efficiency due to beam angle.

As a result, point $S_{1}$ along with the surrounding domain of the Sun sliver base and the width, has the strongest illumination effect on the observation surface. The farther a Sun sliver point is from this domain, the weaker its illumination on the observation surface becomes.

The closest point on the Chromosphere sliver to the observation surface also projects light the most effectively. This point is collinear with the Sun and Moon centers. A similar principle applies to the brightest point on the Corona sliver, which also aligns along the same collinear path. \medskip

$S_{2}$ --- Chromosphere sliver point closest and brightest to points $O^{\prime }$ and $O$

$S_{3}$ --- Corona sliver point closest and brightest to points $O^{\prime }$ and $O$ \medskip

The domains of the Chromosphere and Corona sliver, respectively, surrounding points $S_{2}$ and $S_{3}$, exert the strongest illumination effect on the observation surface.

The observation surface's diameter is several orders of magnitude smaller than the umbra circumference, measuring thousands of times less in length. If a circular arc were sufficiently small compared with the circumference of the full circle, it would appear nearly straight. Likewise, the three sliver domains may be regarded as shafts of light rather than as true arcs.

The three shafts of light project onto the umbra as three corresponding shafts centered around point $O$. \medskip

$S_{1}^{\prime },\,S_{2}^{\prime },\,S_{3}^{\prime }$ --- projections, respectively, of points $S_{1},\,S_{2},\,S_{3}$ onto umbra \medskip

\begin{enumerate}
\item The Sun shaft of light centered on point $S_{1}$ projects onto the shaft centered on point $S_{1}^{\prime }$.
\item The Chromosphere shaft of light centered on point $S_{2}$ projects onto the shaft centered on point $S_{2}^{\prime }$.
\item The Corona shaft of light centered on point $S_{3}$ projects onto the shaft centered on point $S_{3}^{\prime }$.
\end{enumerate}

These shafts dominate illumination on the observation surface. To emphasize their influence in both Figs. 3 and 4, these shafts are thickened. In addition, as totality approaches and the Moon gets close to the Sun edge, all three slivers get shorter as the impact of all three shafts gets pronounced. In total solar eclipses, the Moon disk is greater than the Sun disk. When the size difference is emphasized, all three slivers are somewhat smaller, making the shafts influence even greater.

Although it is known that Earth revolves around the Sun, the Sun appears to move across the sky from the Earth-based perspective. Therefore, the Sun's center motion is depicted with a vector in Fig. 3.

\begin{figure*}[htbp]
\centerline{\includegraphics[width=0.73\textwidth, keepaspectratio]
{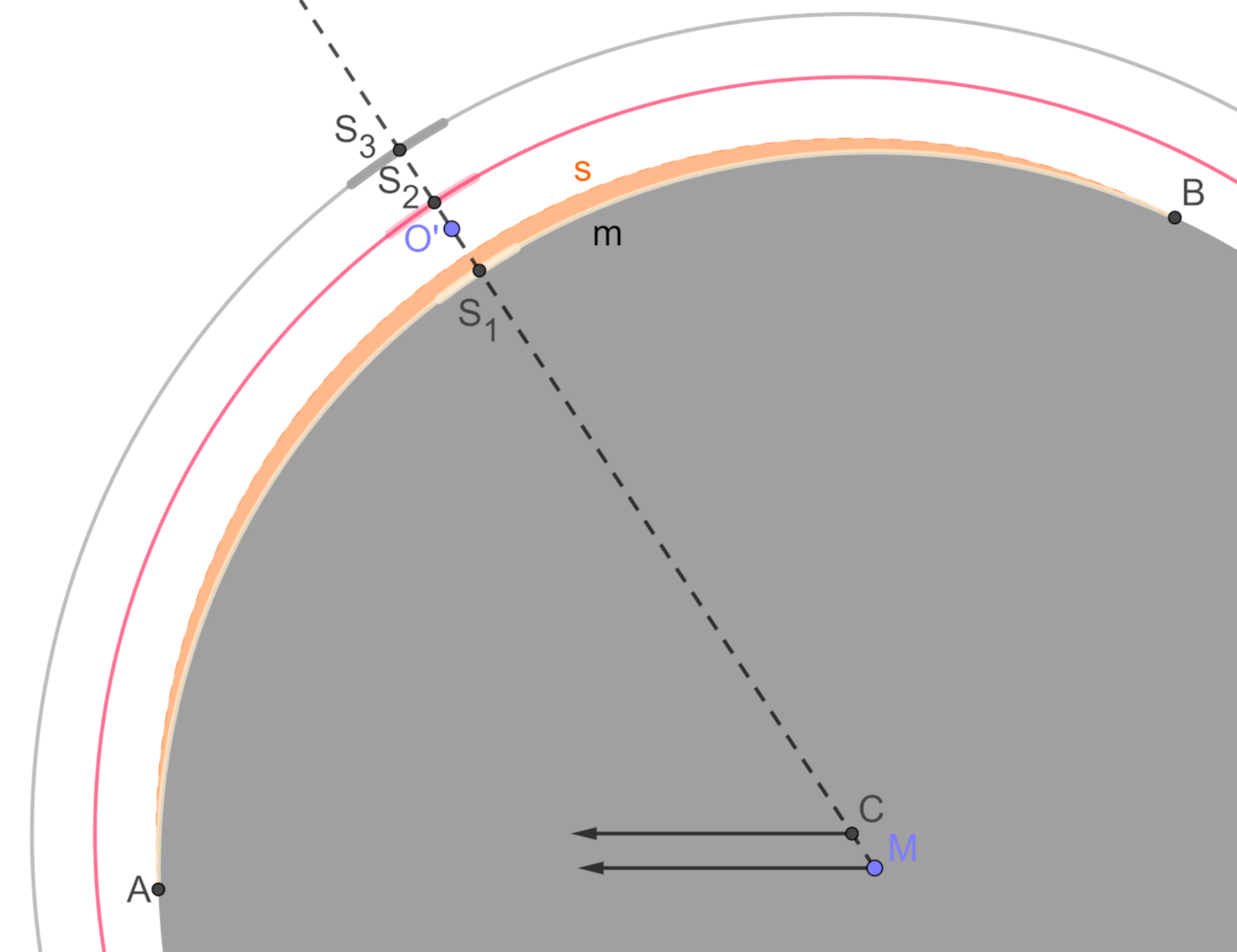}}
\caption{Domains and Shafts}
\end{figure*}

\begin{figure*}[htbp]
\centerline{\includegraphics[width=0.73\textwidth, keepaspectratio]
{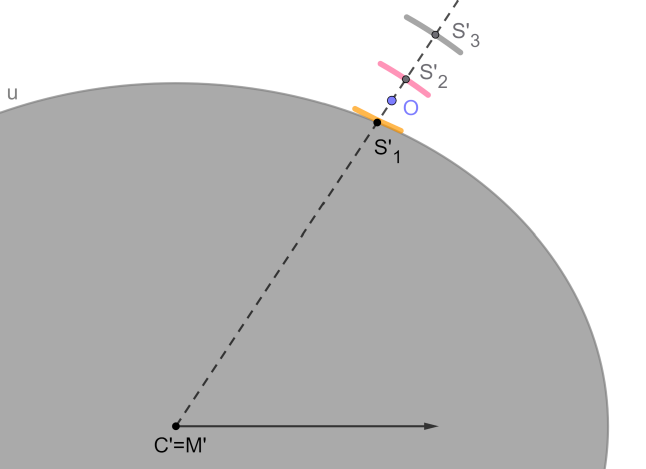}}
\caption{Projection on Observation Surface Plane}
\end{figure*}

\noindent Notice that between the Sun shaft of light and the Chromosphere shaft of light there is the Dark Band shaft of light. And so, there is the Transition Region shaft between the Chromosphere and Corona shafts of light.

However, the Dark Band and Transition Region shafts of light are not graphed in the illustrations (Figs. 3 and 4) due to their small illumination impact. In a more detailed analysis, their contributions can be incorporated into the overall
illumination model, providing a more comprehensive understanding.

It is important to note that all the shafts constructed from the observational point $O$ are positioned tangentially at the appropriate points along the arcs of the circles and the ellipse (Figs. 3 and 4). This means that all three shafts of light on the observation surface are tangential to the Sun and Moon disks, as well as to the umbra.

Let us denote the Sun sliver's width. \medskip

$w$ --- Sun sliver's width

$S$ --- Sun surface point at $w$ \medskip

In Fig. 3 of domains and shafts, we recognize the Sun sliver's width. Point $S$ is omitted from Fig. 3 for clarity but is shown in Fig. 5:
\begin{equation*}
S_{1}S=w
\end{equation*}

Segment $SS_{2}$ is the distance from the Sun surface to the Chromosphere base that we have already estimated:
\begin{equation*}
SS_{2}=\delta _{3}
\end{equation*}

Segment $S_{2}S_{3}$ is the distance from the Chromosphere base to the Corona base that we have also already approximated:
\begin{equation*}
S_{2}S_{3}=d_{1}
\end{equation*}

\begin{case}
This document explores the essential characteristics of shadow bands, assuming the observation surface lies near the umbra centerline.
\end{case}

When the observation surface is positioned along the centerline of the umbra, the Sun sliver's width along with other collinear points are also onto the direction of the centerline (Fig. 5). As a result, the shafts of light of the Sun, the Chromosphere, and the Corona are all perpendicular to the centerline.

\begin{figure}
\centerline{\includegraphics[width=0.7\textwidth, keepaspectratio]
{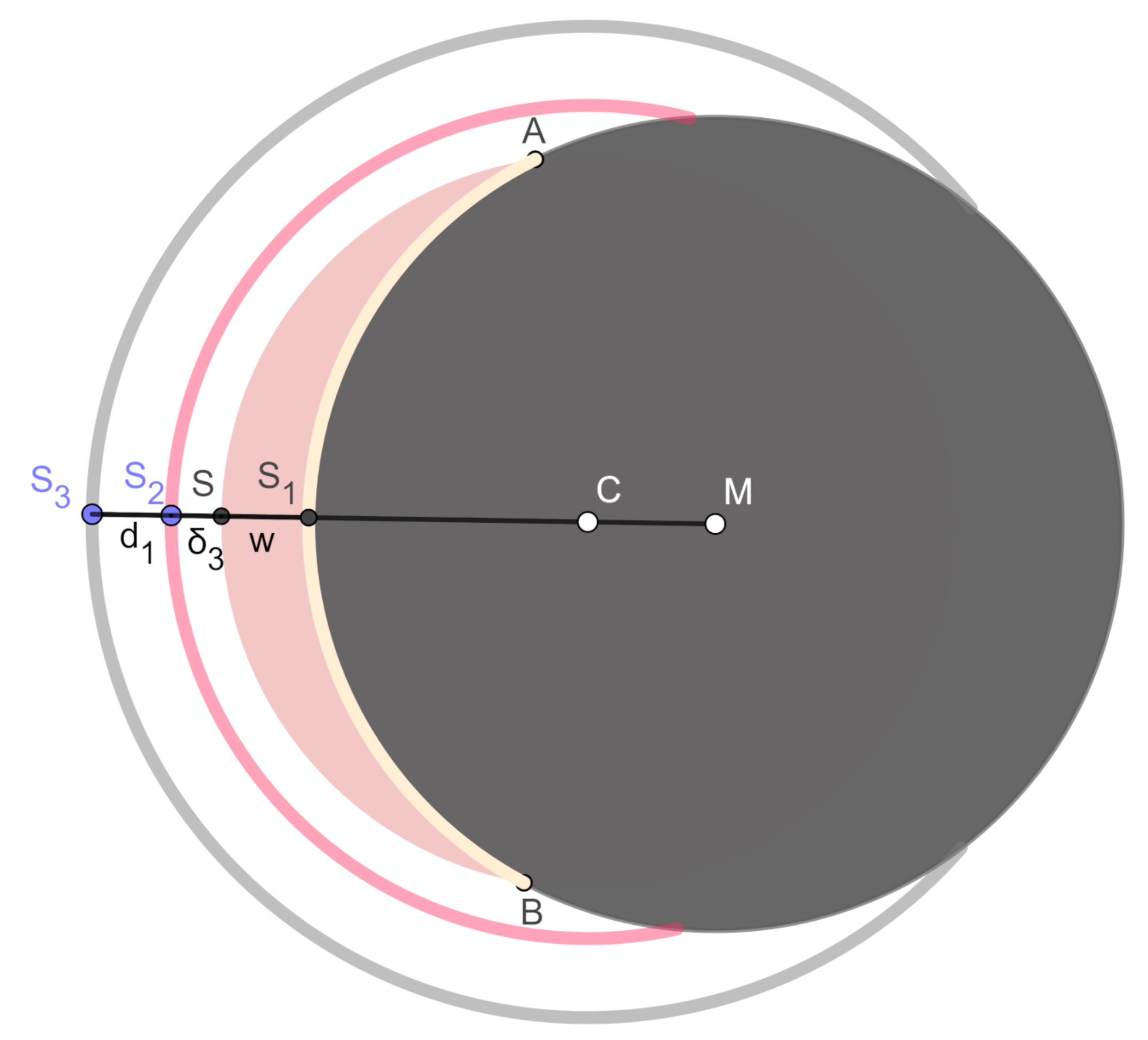}}
\caption{Triple-sliver when Observation Surface at Centerline}
\end{figure}

\noindent As the Sun sliver's width ($w$) shrinks, point $S_{1}$ moves along the Moon's edge and on the verge of totality reaches the Sun surface point $S$.

Shadow bands occur at various moments leading up to totality, typically appearing earlier and later than one minute before, but rarely beyond a few minutes before totality. This makes one minute before totality a reliable reference point. We start by determining the Sun sliver's size exactly one minute before totality. Once established, we then calculate how its width varies at different moments.

Let us find the Sun sliver's width 1 minute before totality:
\begin{equation*}
w=?
\end{equation*}

Since the Sun's diameter is known, determining the Sun sliver's width as a fraction of the diameter will reveal how many times the sliver's width is smaller in comparison. This ratio provides a useful reference for scaling its size relative to the full Sun.

To simplify calculations while maintaining accuracy, we use the fact that the Sun's diameter is relatively small compared to its distance to Earth: \medskip

$z$ --- distance from Sun to Earth
\begin{eqnarray*}
z &=&1.5\times 10^{8}\,\mathrm{km} \\
&=&150\,000\,\mathrm{Mm}
\end{eqnarray*}%
\begin{equation*}
\frac{2R}{z}=\frac{1\,400\,\mathrm{Mm}}{150\,000\,\mathrm{Mm}}=\frac{14}{1\,500}<1\,\%
\end{equation*}

This allows for effective approximations without significant loss of precision.

Over the time interval considered, the relative angular velocity of the Moon with respect to the Sun is assumed to be approximately constant. Consequently, angular displacements are proportional to elapsed time. Because the observation surface lies on the umbral centerline, the motion of the lunar disk across the solar disk may be approximated as linear. Therefore, the fraction of the solar diameter obscured during a given time interval is approximately proportional to the ratio of the corresponding eclipse durations.

The eclipsing of the Sun sliver by the Moon takes one minute, as we have chosen. The eclipsing of the entire Sun begins at C1 (start of eclipse) and ends at C2 (total eclipse), a phase known as the \textit{(first) partial phase}.

Based on statistical observations, the partial phase typically lasts between $70$ and $80$ minutes. To establish a reasonable estimate, we select $75$ minutes, positioned near the midpoint of this relatively narrow range.
\begin{eqnarray*}
\frac{w}{2R} &\approx&\frac{1\,\mathrm{min}}{75\,\mathrm{min}} \\
w &\approx&\frac{2R}{75}
\end{eqnarray*}

We found that one minute before totality the Sun sliver's width is one $75$th of the Sun's diameter. Let us find the percentage.
\begin{equation*}
\frac{1}{75}=\frac{100\,\%}{75}=1.33\,\%
\end{equation*}

The Sun sliver's width fraction of the Sun's diameter, combined with the fraction of the Sun's diameter covered by the Moon, together accounts for the entire Sun's diameter.

The fraction of the Sun's diameter covered by the Moon is known as \textit{eclipse magnitude}---a key measure in describing the extent of the eclipse. \medskip

$\mu$ --- eclipse magnitude (Sun diameter fraction covered by the Moon)

$1-\mu$ --- Sun sliver's width fraction (Sun diameter fraction not covered by the Moon)
\begin{equation*}
\frac{w}{2R}=1-\mu
\end{equation*}

If we were given eclipse magnitude, then we can calculate the Sun sliver's width with the formula
\begin{equation}
w=\left( 1-\mu \right) \cdot 2R  \label{eqWidthEclipseMagnitude}
\end{equation}

Let us find the eclipse magnitude and the Sun sliver's width fraction at various moments below totality.

From the percentage of the Sun sliver’s width $1$ minute before totality, we obtain the corresponding eclipse magnitude:
\begin{eqnarray*}
\mu &=&100\,\%-1.33\,\% \\
&\approx &98.67\,\%
\end{eqnarray*}

At three minutes before totality---when shadow bands may start to appear---the Sun sliver's width is $4\%$ of the Sun's diameter. At this moment, the Moon has covered the entire Sun, except for the darkened limb. This suggests a correlation between limb darkening and shadow band formation.

Having specific time moments before totality can serve as a framework for our analysis. These moments, listed in the first column of Table 1, represent the time measured in seconds or minutes before the C2 event. \medskip

$t$ --- time before totality \medskip

We calculate the eclipse magnitude for each moment from the first column of the table and list it in the second column as a percentage.

To determine the Sun sliver's width values, we apply the formula from Eq. (\ref{eqWidthEclipseMagnitude}) above that substitutes the eclipse magnitudes from the second column and the known Sun's diameter. The Sun sliver's width values are listed in the third column of the table, provided in megameters ($\mathrm{Mm}$) and rounded to two decimal places.

The distance between two bright point sources of light $S_{1}$ and $S_{2}$ is
\begin{equation*}
S_{1}S_{2}=S_{1}S+SS_{2}
\end{equation*}

$d_{3}$ --- distance from Sun shaft to Chromosphere shaft
\begin{equation*}
S_{1}S_{2}=d_{3}
\end{equation*}%
\begin{equation*}
d_{3}=w+\delta _{3}
\end{equation*}

The distance $d_{3}$ is dynamic and is obtained by increasing the width $w$ of the third column with the known distance $\delta _{3}$.
\begin{equation*}
\delta _{3}=0.9\,\mathrm{Mm}
\end{equation*}

We list its values in the fourth column of the table. On the verge of totality, the distance $d_{3}$ shrinks to the distance between the Sun shaft and the Chromosphere shaft, that is $\delta _{3}$.
\begin{equation*}
w=0\implies d_{3}=\delta _{3}
\end{equation*}

The distance between the Corona shaft and the Sun shaft is determined by summing the distances from the Corona shaft to the Chromosphere shaft and from the Chromosphere shaft to the Sun shaft:
\begin{equation*}
S_{1}S_{3}=S_{1}S_{2}+S_{2}S_{3}
\end{equation*}

$d_{2}$ --- distance from Corona shaft to Sun shaft
\begin{equation*}
S_{1}S_{3}=d_{2}
\end{equation*}%
\begin{equation*}
d_{2}=d_{3}+d_{1}
\end{equation*}

The distance $d_{2}$ is dynamic. We calculate the distances by increasing the distance $d_{3}$ from the fourth column with the constant and known distance $d_{1}$.
\begin{equation*}
d_{1}=1.6\,\mathrm{Mm}
\end{equation*}

We list the values in the fifth and final column of the table.

Let us outline our celestial triple-sliver distances (Fig. 6). \medskip

$d_{3}$ --- distance from Sun shaft to Chromosphere shaft

$d_{1}$ --- distance from Chromosphere shaft to Corona shaft

$d_{2}$ --- distance from Corona shaft to Sun shaft \medskip

\begin{table}[htbp]
\centering
\caption{Triple-sliver Distances over Time}
\begin{tabular}{ccccc}
\hline
$t$               & $\mu$       & $w$                  & $d_{3}$              & $d_{2}$              \\ 
\hline
  $0\,\mathrm{s}$ &   $100\,\%$ &     $0\,\mathrm{Mm}$ &  $0.90\,\mathrm{Mm}$ &  $2.50\,\mathrm{Mm}$ \\ 
  $5\,\mathrm{s}$ & $99.89\,\%$ &  $1.54\,\mathrm{Mm}$ &  $2.44\,\mathrm{Mm}$ &  $4.04\,\mathrm{Mm}$ \\ 
 $10\,\mathrm{s}$ & $99.78\,\%$ &  $3.08\,\mathrm{Mm}$ &  $3.98\,\mathrm{Mm}$ &  $5.58\,\mathrm{Mm}$ \\ 
 $20\,\mathrm{s}$ & $99.56\,\%$ &  $6.17\,\mathrm{Mm}$ &  $7.07\,\mathrm{Mm}$ &  $8.67\,\mathrm{Mm}$ \\ 
 $30\,\mathrm{s}$ & $99.33\,\%$ &  $9.25\,\mathrm{Mm}$ & $10.15\,\mathrm{Mm}$ & $11.75\,\mathrm{Mm}$ \\
 $40\,\mathrm{s}$ & $99.11\,\%$ & $12.33\,\mathrm{Mm}$ & $13.23\,\mathrm{Mm}$ & $14.83\,\mathrm{Mm}$ \\
 $50\,\mathrm{s}$ & $98.89\,\%$ & $15.42\,\mathrm{Mm}$ & $16.32\,\mathrm{Mm}$ & $17.92\,\mathrm{Mm}$ \\
$1\,\mathrm{min}$ & $98.67\,\%$ & $18.50\,\mathrm{Mm}$ & $19.40\,\mathrm{Mm}$ & $21.00\,\mathrm{Mm}$ \\
$2\,\mathrm{min}$ & $97.33\,\%$ & $37.00\,\mathrm{Mm}$ & $37.90\,\mathrm{Mm}$ & $39.50\,\mathrm{Mm}$ \\
$3\,\mathrm{min}$ & $96.00\,\%$ & $55.50\,\mathrm{Mm}$ & $56.40\,\mathrm{Mm}$ & $58.00\,\mathrm{Mm}$ \\
$4\,\mathrm{min}$ & $94.67\,\%$ & $74.00\,\mathrm{Mm}$ & $74.90\,\mathrm{Mm}$ & $76.50\,\mathrm{Mm}$ \\
$5\,\mathrm{min}$ & $93.33\,\%$ & $92.50\,\mathrm{Mm}$ & $93.40\,\mathrm{Mm}$ & $95.00\,\mathrm{Mm}$ \\ 
\hline
\end{tabular}
\end{table}

\smallskip

\noindent We may conclude that, to a rough approximation, the Chromosphere and Corona shafts act geometrically as a single shaft up to a dozen seconds before totality.

Fig. 6 is drawn to scale. It illustrates how the distances between the three shafts relate proportionally over 1 minute, 10 seconds, and 5 seconds before totality.
\begin{figure*}[htbp]
\centering
\includegraphics[width=0.95\linewidth]{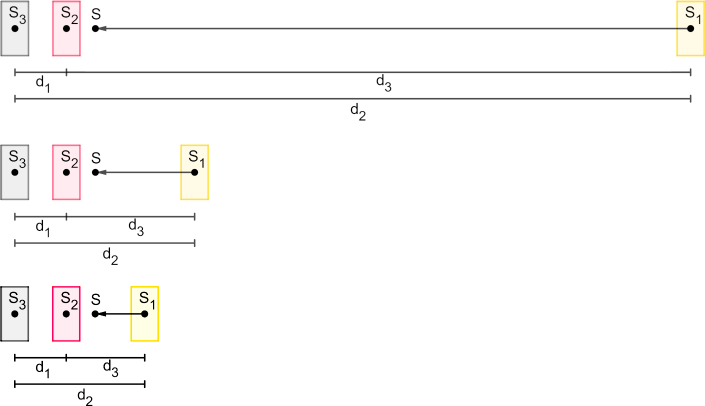}
\caption{Scaled Distances Between Shafts $1\,\mathrm{min}$, $10\,\mathrm{s}$, and $5\,\mathrm{s}$ Before Totality}
\end{figure*}
\noindent At a specific moment during a total solar eclipse, three shafts of light---from the Sun, Chromosphere, and Corona---are geometrically defined by the observer's location (see Figs. 3 and 4).

The Sun shaft of light, directed toward the observation surface, can be interpreted as light passing through one slit in a celestial-scale double-slit ``experiment.'' Similarly, the Chromosphere shaft of light acts as the second slit. The observation surface serves as the screen, or a small portion of it, where interference pattern may emerge in the form of fringes.

The configuration may therefore be interpreted as a celestial analog of Young's double-slit experiment, in which the Sun and Chromosphere slivers act as the effective source pair. For convenience, we refer to this configuration as the \textit{Sun \& Chromosphere-sliver system} or as the \textit{celestial double-sliver system.}

However, this is not a static ``experiment.'' As totality approaches, the Sun sliver---the brightest and most dominant---changes in position and intensity, rendering one of the ``slits'' dynamic. This transforms the analogy into a variable-slit interference model, where fringe behavior evolves in real time.

The Corona sliver, being much dimmer and spatially comparable to the Chromosphere sliver, introduces a third shaft of light. For simplicity, it is excluded from the analysis.

\section{Coherence Analysis}
A star consists of an immense number of emitters, each radiating light at random intervals. Consequently, lights from two separate points on the star's surface are completely uncorrelated---both temporally and spatially incoherent.

This stands in contrast to setups like the double-slit experiment, where coherence is preserved by design. In the stellar case, no fixed phase relationship exists between photons emitted from neighboring regions.

Normal sunlight is broadband, which results in low temporal coherence. However, during an eclipse, this may change due to specific optical effects introduced by the event. To analyze this shift, we will compare the temporal coherence of normal sunlight with that of the Sun \& Chromosphere-sliver.

As starlight propagates over astronomical distances, its spatial coherence evolves. Despite a star's large diameter, its vast remoteness causes it to act, from our viewpoint, as a quasi-point source. Just as stars appear as tiny specks in the night sky, their light exhibits a finite degree of spatial coherence upon arrival at Earth.

During a solar eclipse, the coherence properties of the observed solar radiation may change. As the Moon obscures the Sun, only a narrow sliver of the solar disk remains visible immediately before and after totality. As totality approaches, the angular extent of the emitting region decreases substantially, ultimately vanishing at totality. Since spatial coherence increases as the apparent angular size of a source decreases, the remaining solar sliver is expected to exhibit a greater degree of spatial coherence than the full solar disk.

Now, we examine the temporal and spatial coherence in our celestial double-sliver ``experiment.''

\subsection{Temporal Coherence}
Table 1 shows that, a few minutes before totality, the Moon obscures most of the solar disk, leaving only the solar limb region visible as the lunar disk advances across it.

As noted in the discussion of limb darkening, the solar sliver exhibits both broadband and discrete spectral components. Increased limb darkening suppresses the broadband continuum emission originating from the solar photosphere, thereby enhancing the relative contribution of the discrete spectral features produced in the solar atmosphere.

\begin{enumerate} \renewcommand{\labelenumi}{(\roman{enumi})}
\item The broadband spectrum reduces coherence.
\item The discrete spectrum enhances coherence.
\end{enumerate}

Therefore, limb darkening may increase the degree of temporal coherence in the observed solar radiation. As totality approaches and limb darkening becomes increasingly pronounced, the temporal coherence of the solar sliver is expected to improve.

The Chromosphere spectrum is quasi-monochromatic, making it far more temporally coherent than the broadband spectrum of normal sunlight.

As the Sun sliver narrows, the influence of discrete Chromosphere wavelengths becomes more pronounced on the Sun sliver, leading to the Chromosphere spectral characteristics appearing on both slivers. Since two
light sources sharing the same discrete wavelengths enhance overall temporal coherence, the combined effect of the Sun \& Chromosphere-sliver significantly improves temporal coherence as totality approaches.

On the verge of totality, the combined temporal coherence reaches its maximum.

We now determine the wavelength relevant to our celestial double-sliver experiment.

The strongest spectral line of the Corona ($637.4\,\mathrm{nm}$) is very close to the strongest line of the Chromosphere ($656.3\,\mathrm{nm}$), reinforcing coherence effects. One sliver originates from the Chromosphere, while the other becomes like the Chromosphere on the verge of totality. As a result, the Chromosphere spectrum plays a major role in shaping fringe formation. \medskip

$\lambda $ --- wavelength of celestial double-sliver ``experiment'' 
\begin{eqnarray*}
\lambda &\approx &656.3\,\mathrm{nm} \\
&\approx &637.4\,\mathrm{nm} \\
&\approx &650\,\mathrm{nm}
\end{eqnarray*}

Meanwhile, the Dark Band remains significantly dim compared to these two approximately equal dominant spectral lines.

\subsection{Spatial Coherence}
As totality approaches, the Sun sliver's width---and consequently its angular width---diminishes significantly. Three minutes before totality, our calculations show that the Sun sliver's width is about $4\%$ of the Sun's diameter; one minute before totality, it narrows farther to $1.33\%$. On the verge of totality, the sliver shrinks to none. This dramatic reduction greatly enhances spatial coherence, as the Sun's angular size is large in comparison to the progressively smaller sliver's widths.

Thus, after a certain moment leading up to totality, the Sun shaft of light is expected to exhibit enhanced spatial coherence. Even a few minutes before totality, it should maintain at least partial coherence, given the tiny remaining sliver.

Additionally, the greatest illumination effect arises from the Sun sliver base and its width, which constitutes only a small distance portion of the crescent width.

The bright Chromosphere (excluding the Dark Band) can extend up to $2\,000\,\mathrm{km}$, although it is typically thinner. Now, we estimate its ratio relative to the Sun's diameter:
\begin{equation*}
\frac{2\,000\,\mathrm{km}}{2R}=\frac{2\,\mathrm{Mm}}{2\cdot 700\,\mathrm{Mm}}
=\frac{1}{700}=\frac{1}{7}\cdot\frac{1}{100}<0.2\,\%
\end{equation*}

Our calculations show that one minute before totality, the Chromosphere sliver's width is less than $0.2\%$ of the Sun's diameter. Owing to the dynamic nature of the Chromosphere, however, this width, and consequently its angular extent, may be several times smaller. Hence, the Chromosphere sliver is expected to exhibit a greater degree of spatial coherence than the full solar disk.

For a double-slit experiment to produce a clear interference pattern, the slits must satisfy certain requirements.

The slits must be sufficiently narrow for the transmitted light to diffract and spread outward, allowing the wavefronts emerging from the two slits to overlap. The superposition of these wavefronts produces the interference pattern observed as fringes on the screen.

However, the slits cannot be arbitrarily wide. As the slit width increases, diffraction weakens, reducing the overlap between the emerging wavefronts and diminishing the visibility of the interference fringes.

As totality approaches, the overall brightness decreases dramatically. At this stage, the principal sources of illumination originate from the three solar slivers and, more remarkably, from the Sun and Chromosphere shafts of light.

Each sliver can be observed from a wide range of viewing directions because the Sun and its atmospheric layers emit and scatter light in many directions. As a result, light from different slivers can overlap at the observer's location, creating the conditions necessary for interference.

In the conventional double-slit experiment, the slits diffract the incident light, causing it to spread and allowing the emerging wavefronts to overlap. However, in the present case, the solar slivers act as naturally extended light sources whose radiation already propagates over a range of directions. Consequently, no additional diffraction mechanism is required to produce the overlap of light originating from different slivers.

Nevertheless, the slivers must remain sufficiently narrow. If a sliver is too wide, different parts of it contribute light arriving at different angles, reducing the visibility of the resulting interference pattern. Therefore, the slivers should be narrow enough to approximate line or point-like sources in the transverse direction.

We now compare the slit width in a conventional double-slit experiment with the width of the Chromosphere sliver in the proposed double-sliver configuration. In a typical laboratory double-slit experiment, the distance between the double-slit and the observation screen is of the order of $1$ to $3\,\mathrm{m}$. For the present comparison, we adopt a representative distance of $z_y=2\,\mathrm{m}$. The width of the slit usually lies in the range of 10$\,\mu\mathrm{m}$ to 50$\,\mu\mathrm{m}$. 

The distance from the Sun \& Chromosphere-sliver to the observation surface can be interpreted in Young's double-slit experiment as the distance from the double-slit to the screen. To compare celestial double-sliver ``experiment'' and Young's double-slit experiment, we will scale down the distance between double-sliver and observation surface to the distance between double-slit and wall. We already know the distances, so the ratio is
\begin{equation*}
\frac{z_{y}}{z}=\frac{2\,\mathrm{m}}{150\,000\,\mathrm{Mm}}
\end{equation*}%

Now let us dilate the Chromosphere sliver's width and then compare it with the slit width in Young's double-slit experiment:
\begin{eqnarray*}
\frac{2\,\mathrm{m}}{150\,000\,\mathrm{Mm}}\cdot 2.5\,\mathrm{Mm} &=&\frac{5\,%
\mathrm{m}}{150\,000\,} \\
&=&\frac{5\,000\,000\,\mathrm{\mu m}}{150\,000\,} \\
&=&\frac{500\,}{15\,}\,\mathrm{\mu m} \\
&=&33\,\mathrm{\mu m}
\end{eqnarray*}

The value obtained falls within the range of slit widths commonly used in Young's double-slit experiments, namely 10$\,\mu\mathrm{m}$ to 50$\,\mu\mathrm{m}$. Moreover, the base of the Chromosphere sliver, which constitutes the brightest region of the Chromosphere, may be considerably narrower. This comparison indicates that the geometric scale of the Chromosphere sliver is comparable to the slit widths commonly employed in Young-type interference experiments.

Because these naturally formed slits have a heterogeneous origin, the coherence of the emitted light is limited. Hence, the resulting interference pattern cannot match the sharpness and contrast typically achieved in laser double-slit experiments.

We now compare the slit separation in a conventional double-slit experiment with the separation between the chromospheric light shaft and the solar light shaft five minutes before totality. In a successful double-slit experiment, the slit separation typically ranges from fractions of a millimeter to a few millimeters, depending on the experimental configuration. For a representative slit-to-screen distance of $z_y=2\,\mathrm{m}$, the slit separation is commonly of the order of a few millimeters.

Now we scale the separation between the Chromosphere sliver and the Sun sliver base five minutes before totality, as given in Table 1 ($d_3=93\,\mathrm{Mm}$), and then compare it with the slit separation in Young's double-slit experiment:
\begin{eqnarray*}
\frac{2\,\mathrm{m}}{150\,000\,\mathrm{Mm}}\cdot 93\,\mathrm{Mm} &=&\frac{2}{%
150\,000\,}\cdot 93\,\mathrm{m} \\
&=&\frac{186}{150\,}\cdot 10^{-3}\,\mathrm{m} \\
&\approx &1\,\mathrm{mm}
\end{eqnarray*}

The obtained value is of the order of millimeters, comparable to slit separations commonly used in Young's double-slit experiments. This comparison indicates that the geometric proportions of the celestial double-sliver system are similar to those encountered in laboratory interference configurations about 5 minutes before totality. As totality approaches, the separation between the shafts decreases, resulting in larger predicted fringe widths and spacings. Simultaneously, the narrowing solar sliver reduces the apparent angular extent of the emitting region, which is expected to enhance spatial coherence. Consequently, spatial coherence is expected to reach its maximum immediately before totality.

The scaling preserves the angular dimensions of the system and therefore preserves the relative geometric relationships between source size, source separation, and observation distance. Since spatial coherence depends strongly on the apparent angular extent of the emitting regions, this comparison provides additional support for the analogy between the celestial double-sliver system and laboratory interference configurations.

All the preceding observations support the conclusion that the temporal and spatial coherence is significantly enhanced when the Sun sliver is within the Sun limb region, occurring a few minutes before totality. The combined potential of the Sun \& Chromosphere-sliver, and especially the shafts of light, to generate interference fringes is substantially greater than that of the entire Sun.

As totality approaches, the temporal and spatial coherence of the solar and chromospheric slivers is expected to increase, reaching its highest level immediately before totality. This suggests that interference between light originating from the chromospheric sliver and the solar sliver may occur, producing alternating bright and dark fringes analogous to those observed in a double-slit experiment.

We now estimate the range of fringe widths and fringe spacings expected in the double-sliver configuration. For this purpose, the Van Cittert-Zernike theorem provides a useful theoretical framework.

According to the Van Cittert-Zernike theorem, reducing the apparent angular size of an incoherent source increases the spatial coherence of the received field \cite{Born2017}. For a uniform circular source subtending an angular diameter $\theta$ at the observer and observed at wavelength $\lambda$, the complex degree of coherence first vanishes at a transverse separation approximately equal to the spatial coherence length
\begin{eqnarray*}
\ell _{c}\approx1.22\frac{\lambda}{\theta}
\end{eqnarray*}

We now consider a circular source that encompasses both the chromospheric light shaft and the solar light shaft 5 seconds before totality. The diameter of this region is taken to be the sum of the chromospheric altitude above its base and the width of the solar sliver at that time. We estimate the former as $1\,500\,\mathrm{km}=1.5\,\mathrm{Mm}$, while Table 1 gives a solar-sliver width of approximately $1.5\,\mathrm{Mm}$. This yields a total source diameter of approximately $3\,\mathrm{Mm}$. To evaluate the angular diameter $\theta$, we apply the small-angle approximation:
\begin{eqnarray*}
\theta \approx \frac{3\,\mathrm{Mm}}{150\,000\mathrm{Mm}}
\end{eqnarray*}

Therefore,
\begin{eqnarray*}
\frac{1}{\theta } &\approx &\frac{150\,000\,\mathrm{Mm}}{3\,\mathrm{Mm}} \\
&\approx &50\,000
\end{eqnarray*}%

The spatial coherence length is
\begin{eqnarray*}
\ell _{c} &\approx &1.22\cdot \frac{1}{\theta }\cdot \lambda  \\
&\approx &1.22\cdot 50\,000\cdot 650\,\mathrm{nm} \\
&\approx &6\cdot 10^{4}\cdot 6.5\cdot 10^{2}\cdot 10^{-9}\,\mathrm{m} \\
&\approx &40\cdot 10^{-3}\,\mathrm{m} \\
&\approx &40\,\mathrm{mm} \\
&\approx &4\,\mathrm{cm}
\end{eqnarray*}

The source diameter is chosen to encompass both the solar and chromospheric shafts. The resulting coherence length therefore represents a characteristic scale for their mutual coherence and provides an estimate of the largest interference structures that may remain observable.

The coherence analysis suggests that interference structures on centimeter scales are plausible.

Notably, the characteristic coherence length derived above and the independently predicted fringe width and spacing obtained in the next section are both on the order of a few centimeters.

\section{Predictions}
Given that the Sun \& Chromosphere-sliver is expected to generate fringes, a key question arises regarding fringe width and spacing.

The experimental results in the proceedings \emph{Spatial coherence of sunlight: first direct measurement} confirm previously established theoretical predictions regarding the spatial coherence of sunlight \cite{Mashaal2012}. The theoretical prediction is that for an average wavelength of $510\,\mathrm{nm}$ the spatial coherence length of sunlight is approximately $50\,\mathrm{\mu m}$ at the Earth's surface under clear skies. However, if the Earth's atmosphere causes cloudy and foggy weather, the spatial coherence length may decrease even a dozen times.

If the Moon or another object were to obscure the Sun, leaving only two diametrically opposite regions of the solar disk visible, those regions would act as two widely separated coherent source areas. In monochromatic light near the peak solar wavelength ($510\,\mathrm{nm}$) , they would generate an interference pattern whose fringe spacing would be approximately $50\,\mathrm{\mu m}$, in agreement with theoretical predictions and experimental measurements.

In our double-sliver ``experiment,'' the distance between the Sun and Chromosphere slivers, observed in quasi-monochromatic light five seconds before totality, is estimated from Table 1 to be
\begin{equation*}
d_{3}=2.4\,\mathrm{Mm}
\end{equation*}

Compared to the Sun's diameter, the distance between slivers is smaller by a factor evaluated as follows:
\begin{equation*}
\frac{2R}{d_{3}}=\frac{2\cdot 700\,\mathrm{Mm}}{2.4\,\mathrm{Mm}}=\frac{700}{1.2}\approx 600
\end{equation*}

Therefore, the fringe width and spacing in our double-sliver ``experiment'' are $600$ times larger than $50\,\mathrm{\mu m}$, owing to the inverse proportionality between slit separation and fringe width (or spacing).

Therefore, the fringe width and spacing for the average solar wavelength are
\begin{eqnarray*}
600\cdot 50\,\mathrm{\mu m} &=&30\cdot 1\,000\,\mathrm{\mu m} \\
&=&30\,\mathrm{mm} \\
&=&3\,\mathrm{cm}
\end{eqnarray*}

To refine this further, we will also account for the effect of wavelength on our calculations.

Due to the proportionality with wavelength, the fringe width and spacing can be calculated as follows:
\begin{eqnarray*}
\frac{650\,\mathrm{nm}}{510\,\mathrm{nm}}\cdot 3\,\mathrm{cm} &=&\frac{65}{51}\cdot 3%
\,\mathrm{cm} \\
&\approx &\frac{68}{17}\,\mathrm{cm} \\
&\approx &4\,\mathrm{cm}
\end{eqnarray*}

The formula for the width and spacing of the fringe yields almost the same result: \medskip

$z$ --- distance from slits to screen

$d$ --- distance between slits

\begin{eqnarray*}
\frac{\lambda z}{d} &=&\frac{650\,\mathrm{nm}\cdot 150\,000\,\mathrm{Mm}}{2.4\,\mathrm{Mm}} \\
&=&\frac{6.5\cdot 10^{2}\cdot 10^{-9}\,\mathrm{m}\cdot 1.5\cdot 10^{5}}{2.4} \\
&=&\frac{6.5\cdot 1.5}{2.4}\cdot 10^{-2}\,\mathrm{m} \\
&\approx &4\,\mathrm{cm}
\end{eqnarray*}

On the verge of totality, Table 1 gives a sliver separation of $0.9\,\mathrm{Mm}$, which is more than a factor of two smaller than the separation 5 seconds before totality. Since fringe width and fringe spacing are inversely proportional to the separation of the slivers, both quantities increase by more than a factor of two, reaching approximately $10\,\mathrm{cm}$.

However, the altitudes of the Chromosphere and the Dark Band are highly variable, so these values should only be regarded as rough estimates.

\begin{conclusion}
A few seconds before totality, the fringe width and spacing above Earth's atmosphere, or on the ground under clear skies, are estimated to be of the order of a few centimeters. However, the highly variable altitudes of the Chromosphere and the Dark Band may cause these values to be several times larger or smaller.
\end{conclusion}

This estimation of the fringe width and spacing is closely aligned with the shadow band width and spacing.

The distance between the Sun shaft and the Chromosphere shaft ($d_3$), decreases as totality approaches. Since fringe spacing is inversely proportional to source separation, the corresponding fringe width and spacing increase with time. This trend is consistent with the variation in shadow-band spacing reported by J. L. Codona and discussed in Section ``Width and Spacing.''

The fringes are generated at altitudes well above Earth's atmosphere, where they can be observed and detected. The Earth's atmosphere preferentially absorbs shorter wavelengths of light, particularly in the ultraviolet and blue regions of the spectrum.

As a consequence of atmospheric scattering and absorption, the effective mean wavelength of sunlight above the atmosphere is shorter than that observed at the Earth's surface, where longer wavelengths are transmitted more efficiently. Since fringe width and spacing are directly proportional to wavelength, the corresponding fringe dimensions above the atmosphere are expected to be somewhat smaller. This trend is consistent with the high-altitude balloon observations of shadow bands discussed in Section ``Observational Constraints.''

From the geometric model, we infer that the interference fringes appear locally as lines tangent to the umbra boundary. Consequently, they form a globally concentric pattern around the umbra.

The concentric fringe pattern produces observational effects that depend on the observer's position relative to the umbral centerline. Near the centerline, the fringes appear nearly parallel, whereas farther from the centerline they exhibit more pronounced changes in orientation. In addition, the fringes may remain visible for a longer time because the umbral center does not approach the observer directly. Both predictions are supported by observational reports of shadow bands.

These fringes propagate along the umbra at speeds far too high to be effectively captured by conventional photographic cameras. The rapid motion of the fringes may limit the ability of visual observations and conventional imaging systems to accurately determine their intrinsic properties. Any visual observation or video recording of fringes may differ significantly from their true width and spacing, speed, direction, and other characteristics.

Hence, we cannot rely on a single observation, but instead we should make multiple observations and use statistical and other scientific methods to make empirical conclusions more reliable.

Near totality, the Moon's irregular topography shapes the solar sliver into a zigzag pattern. This effect manifests itself in the form of Baily's beads. In a conventional double-slit experiment, both slits are straight and parallel, producing a regular interference pattern with evenly spaced bright and dark fringes. However, if one slit remains straight while the other is rugged, the symmetry of the system is broken, altering the resulting interference field. Consequently, the interference fringes are also expected to exhibit a zigzag structure. This effect becomes increasingly pronounced as totality approaches. This behavior is consistent with reports that shadow bands can show a snake-like or undulating appearance.

The other observational properties of shadow bands not discussed in this article can be explained within the proposed framework without significant additional complexity.

\section{Conclusions and Discussion}
When we restate all dark fringe distinct properties emerging from the celestial triple-sliver light interference pattern, we recognize that they match those of shadow bands: their appearance, size, speed, direction, duration, and more. The predicted properties of the proposed interference pattern are consistent with the reported properties of shadow bands, supporting the interpretation that the two phenomena may share a common origin.

Therefore, any conclusions we have drawn about dark fringes apply directly to shadow bands.

From this point forward, we will avoid referring to them as ``dark fringes'' or ``hypothetical shadow bands.'' Instead, we will use the unified scientific term: shadow bands.

Recall that shadow bands are not exclusive to total solar eclipses. While our focus has been on the Sun as the source, other stars---provided they exhibit sufficient light coherence---could also produce them.

For this to occur, such stars must possess bright ring-like structures and intervening dark layers within their atmospheres. These features would mirror the Sun's architecture: the luminous bases of the Chromosphere and Corona, separated by the obscuring strata of the Dark Band and the Transition Region.
\begin{definition}
\textbf{Shadow bands} are dark fringes that may appear on an observation surface when a star's disk is nearly obscured. They arise from the star's dark limb zone and at least one bright sliver-like atmospheric layer, flanked by adjacent dark regions.\\
This configuration primarily produces interference between light originating from the brightest regions near the sliver centers, which act as localized light sources, analogous to the slits in a classical multiple-slit experiment.
\end{definition}

\begin{definition}
\textbf{Solar shadow bands} mostly emerge due to the light interference from the shafts of the Sun, Chromosphere, and Corona-slivers---a celestial analog of a triple-slit experiment---activated when the Sun nearly eclipses.\\
These luminous regions are embedded within the darker surroundings of the Dark Band and the Transition Region.
\end{definition}

\begin{corollary}
Because the Corona base is relatively dim and closely aligned with the Chromosphere base, the resulting interference may be approximated as a Sun \& Chromosphere-sliver pattern.\\
The celestial double-sliver configuration may be regarded as a natural, time-dependent analog of Young's double-slit experiment, with the slit separation and illumination conditions evolving as totality approaches.
\end{corollary}

Without the Dark Band, there would be no two distinct slivers and, consequently, no mutual interference between them capable of producing shadow bands. The central conclusion regarding the formation of solar shadow bands is summarized below:
\begin{conclusion}
Dark bands of a solar eclipse are caused by the Dark Band.    
\end{conclusion}

The appearance of shadow bands, including their width and spacing, is governed by a complex interplay of factors, including the following:

\begin{enumerate} \renewcommand{\labelenumi}{(\roman{enumi})}
\item the changing geometry of the lunar occultation of the Sun,
\item the width and dynamics of solar layers near the limb,
\item the degree of partial light coherence in the incoming light,
\item the structural features of the umbra,
\item the conditions of Earth's atmosphere, such as turbulence and refractive gradients, and
\item surface reflectivity.
\end{enumerate}

As a result, shadow bands exhibit highly dynamic and inherently unpredictable behavior, varying not only from eclipse to eclipse, but even from moment to moment within a single event.

Additional material relevant to these conclusions---including supplementary observational data and extended derivations and analysis---is available in the author's book \emph{The Optics of Shadow Bands} \cite{MathallicA2025p}.

\begin{backmatter}



\bmsection{Disclosures}
The author declares no conflicts of interest.

\end{backmatter}


\bibliography{sample}

@Book{BAA1906,
  author     = {{British Astronomical Association}},
  publisher  = {{British Astronomical Association}},
  title      = {The total solar eclipse 1905: Reports of observations made by members of the {B}ritish {A}stronomical {A}ssociation},
  year       = {1906},
  address    = {London},
  note       = {Observational report of the August 30, 1905 total solar eclipse. Public domain},
  pages      = {49--52},
  url        = {https://archive.org/details/totalsolareclips00britrich}
}

@Book{Born2017,
  author    = {Born, Max and Wolf, Emil},
  publisher = {Cambridge University Press},
  title     = {Principles of optics},
  year      = {2017},
  address   = {Cambridge [u.a.]},
  edition   = {Seventh (expanded) edition, 13th printing. edition},
  isbn      = {9780521642224},
  pagetotal = {952},
  ppn_gvk   = {1638168520},
  subtitle  = {Electromagnetic theory of propagation, interference and diffraction of light},
}

@Article{Codona1986,
  author        = {Johanan L. Codona},
  journal       = {Astronomy and Astrophysics},
  title         = {The scintillation theory of eclipse shadow bands},
  year          = {1986},
  issn          = {0004-6361},
  month         = aug,
  note          = {Please note that the article is copyrighted and may not be freely reproduced},
  number        = {2},
  pages         = {415--427},
  volume        = {164},
  publisher     = {EDP Sciences},
  url           = {https://articles.adsabs.harvard.edu/full/1986A&A...164..415C}
  }

@Article{Codona1991,
  author     = {Johanan L. Codona},
  journal    = {Sky and Telescope},
  title      = {The enigma of shadow bands},
  year       = {1991},
  issn       = {0037-6604},
  month      = may,
  number     = {5},
  pages      = {482--487},
  volume     = {81}
}

@Article{Feldman1938,
  author        = {Richard L. Feldman},
  journal       = {Popular Astronomy},
  title         = {Shadow bands -- {P}art {I}},
  year          = {1938},
  month         = apr,
  note          = {This article is believed to be in the public domain},
  pages         = {187--200},
  volume        = {46},
  url           = {https://adsabs.harvard.edu/full/1938PA.....46..187F}
}

@Article{Gaviola1948,
  author     = {E. Gaviola},
  journal    = {Popular Astronomy},
  title      = {On shadow bands at total eclipses of the sun},
  year       = {1948},
  note       = {This article is believed to be in the public domain},
  pages      = {353--359},
  volume     = {56},
  url        = {https://ui.adsabs.harvard.edu/scan/manifest/1948PA.....56..353G}
}

@Article{Henry1906,
  author     = {A. J. Henry},
  journal    = {Monthly Weather Review},
  title      = {Observations of `shadow bands' without an eclipse},
  year       = {1906},
  month      = may,
  number     = {5},
  pages      = {227},
  volume     = {34}
}

@Article{Hults1971,
  author        = {Hults, Malcolm E. and Burgess, Roger D. and Mitchell, Daniel A. and Warn, Duane W.},
  journal       = {Nature},
  title         = {Visual, photographic and photoelectric detection of shadow bands at the {M}arch 7, 1970, solar eclipse},
  year          = {1971},
  issn          = {1476-4687},
  month         = may,
  number        = {5300},
  pages         = {255--258},
  volume        = {231},
  doi           = {10.1038/231255a0},
  publisher     = {Springer Science and Business Media LLC}
}

@Article{Madhani2020,
  author        = {Madhani, Janvi P. and Chu, Grace E. and Gomez, Carlos Vazquez and Bartel, Sinjon and et al.},
  journal       = {Journal of Atmospheric and Solar-Terrestrial Physics},
  title         = {Observation of eclipse shadow bands using high altitude balloon and ground-based photodiode arrays},
  year          = {2020},
  issn          = {1364-6826},
  month         = dec,
  note          = {Preprint available on arXiv site (not in the public domain)},
  number        = {105420},
  volume        = {211},
  doi           = {10.1016/j.jastp.2020.105420},
  eprinttype    = {arxiv},
  publisher     = {Elsevier BV}
}

@InProceedings{Mashaal2012,
  author        = {Mashaal, Heylal and Goldstein, Alex and Feuermann, Daniel and Gordon, Jeffrey M.},
  booktitle     = {Nonimaging Optics: Efficient Design for Illumination and Solar Concentration IX},
  title         = {Spatial coherence of sunlight: first direct measurement},
  year          = {2012},
  editor        = {Winston, Roland and Gordon, Jeffrey M.},
  month         = oct,
  pages         = {84850A},
  publisher     = {SPIE},
  volume        = {8485},
  doi           = {10.1117/12.928449},
  issn          = {0277-786X},
  url           = {https://ui.adsabs.harvard.edu/abs/2012SPIE.8485E..0AM/abstract}
}

@Book{MathallicA2025p,
  author    = {MathallicA},
  publisher = {MathallicA},
  title     = {The optics of shadow bands: Causation and effect},
  year      = {2025},
  edition   = {First},
  isbn      = {978-86-908194-0-9},
  month     = sep,
  note      = {Paperback edition},
  url       = {https://www.amazon.com/dp/8690819401/}
}

@Article{OMeara2009,
  author     = {Stephen James O'Meara},
  journal    = {Astronomy},
  title      = {Secret sky -- {S}earching for shadow bands},
  year       = {2009},
  issn       = {0091-6358},
  month      = feb,
  number     = {4},
  pages      = {18--19},
  volume     = {37},
  url        = {https://www.astronomy.com/observing/stephen-james-omearas-secret-sky-searching-for-shadow-bands/}
}
\end{document}